\documentclass[twocolumn,amsmath,amssymb,floatfix,superscriptaddress,prb]{revtex4-2}

\usepackage{bm,graphicx,url,epsf,color}
\usepackage{amssymb}
\usepackage{amsthm}
\usepackage[colorlinks=true,linkcolor=blue,citecolor=blue]{hyperref}
\usepackage{comment}

\definecolor{purple}{rgb}{0.5,0,0.6}
\usepackage{ulem}
\definecolor{darkblue}{rgb}{0,0,0.5}
\definecolor{darkgreen}{rgb}{0,0.5,0}
\definecolor{darkred}{rgb}{.7,0,0}
\definecolor{purple}{rgb}{0.5,0,0.6}
\definecolor{orange}{rgb}{1,0.5,0}
\definecolor{grey}{rgb}{.6,.6,.6}
\definecolor{lightpink}{rgb}{1,0.7,0.75}
\definecolor{pink}{rgb}{1,0.4,0.58}
\definecolor{deeppink}{rgb}{1,0.08,0.58}

\renewcommand{\emph}[1]{\textit{#1}}

\newcommand{\ngo}{N_{\rm 1, g}}
\newcommand{\ngt}{N_{\rm 2, g}}
\newcommand{\tc}{\tau_{\rm c}}
\begin{document}
\title{Two-sites quantum island in the quasi-ballistic regime}

\author{Deepak B. Karki}
\affiliation{Division of Quantum State of Matter, Beijing Academy of Quantum Information Sciences, Beijing 100193, China}
\author{Edouard Boulat}
\affiliation{Universit\'e  de  Paris, CNRS, Laboratoire  Mat\'eriaux  et  Ph\'enom\`enes  Quantiques, 75013  Paris,  France}
\author{Christophe Mora}
\affiliation{Universit\'e  de  Paris, CNRS, Laboratoire  Mat\'eriaux  et  Ph\'enom\`enes  Quantiques, 75013  Paris,  France}

\begin{abstract}
Quantum Hall edge channels can be combined with metallic regions to fractionalize electrons and form correlated impurity models. We study a minimal device, that has been experimentally achieved quite recently, with two floating islands connected to three edge channels via quantum point contacts in the integer quantum Hall regime. At high transparency of the quantum point contacts, we establish a mapping to the boundary sine-Gordon model and thereby reveal the nature of the quantum critical point. We deduce from this mapping universal expressions for the conductance and noise, in agreement with the experimental findings, and discuss the competition between Kondo-like screening of each individual island and the cooperative transfer of electrons between them. We further predict that the device operated at finite voltage bias produces fractional charges $e^* =e/3$ and propose a generalization to $N$ islands with the fractional charge $e^* =e/(N+1)$.
\end{abstract}

\maketitle

\section{Introduction}

Models of quantum impurities embedded in fermionic environments entail a rich zoology of phase transitions with non-Fermi liquid scaling~\cite{Wiegmann_1983,Hewson, Coleman_book, sachdev_2011}. These models underpin many properties of correlated materials such as heavy fermions~\cite{Gegenwart_2008}. They can also be realized in superconducting~\cite{Martinez2019,Kaur2021,Roy2019} or semiconducting nanostructures with a versatile control on the parameters driving the transitions. For example, the Kondo screening of local spin degrees of freedom  is routinely observed in semiconducting quantum dots in two-dimensional electron gases~\cite{Goldhaber_nat(391)_1998, Cronenwett_1998}, carbon nanotubes~\cite{Herrero_2005} or nanowire devices~\cite{Jesper_2006}.
The screening of the impurity spin by a second channel~\cite{Mitchell_2012a,Sela_2016} is however difficult, but not impossible, to achieve experimentally with quantum dots~\cite{Yuval_2003, Potok_NAT(446)_2007, Mebrahtu_2012, Keller_2015}.

An alternative route for Kondo screening with more than one channel is offered by the charge Kondo effect~\cite{Matveev_1991, Flensberg_1993,Matveev1995, Furusaki1995b,LeHur_2002,mora2013}. In the charge version, two quasi-degenerate and discrete charge states of a metallic island are screened by the tunnel coupling to a lead. It has been realized in an experiment~\cite{Pierre_2015, Pierre_2016,Pierre_2018, Pierre_2018a, Pierre_2019} coupling a micron-size floating island to the edge channels of a two-dimensional electron gas tuned in the integer quantum Hall regime. The transmission to the metallic island is controlled by nearby quantum point contacts (QPC) playing the role of tunnel junctions. Kondo scalings, crossovers and fixed points have thus been measured with unprecedented control and detail in the two- and three-channel cases, together with the renormalization-group relevant channel asymmetries~\cite{Pierre_2015, Pierre_2018}. Interestingly, the gate-controlled quantum point contacts can be tuned to be almost ballistic, {\it i.e.} with a large transparency, in which case the temperature or voltage scalings no longer emulate a Kondo model since many charge states are involved. Despite the absence of a Kondo mapping in the quasi-ballistic limit, the zero-temperature quantum critical point is continuously connected to the one at small transparencies where Kondo scaling holds. This continuity argument preserves the properties at zero energy: the fractional entropy, the leading temperature/voltage scaling and more generally the operator content of the zero-temperature quantum critical point are the same regardless of the transparencies. Only the temperature/voltage evolution of observables, such as the differential conductance, are different and depend on the transparencies, with a Kondo mapping restricted to low transparency.

Recently, a two-site version of the charge Kondo effect has been realized with two coupled metallic islands in a two-dimensional electron gas in a GaAs/AlGaAs heterostructure, and a novel zero-temperature quantum critical point has been identified by comparing experimental data with numerical renormalization group (NRG) calculations~\cite{pouse2021}. The competition between the screening of each individual island charge and the mediated charge coupling between the two islands may offer an insight on the competition between Kondo screening and collective magnetic ordering in correlated materials. In the case of an exchange coupling between two spins, the model has been coined as the two-impurity Kondo model~\cite{Jayaprakash_1981,Affleck_1992,Sela_2011,Mitchell_2012b} and discussed generally in the context of quantum dots~\cite{Craig_2004,Simon_2005,Vavilov_2005,Zarand_2006}.

In this paper, we consider the geometry of this experiment in the quasi-ballistic regime where each QPC is set close to full transparency. We recover analytically the properties of the zero-temperature quantum critical point observed experimentally and in NRG calculations which, as argued above, are universal in the sense that they do not depend on the transparencies of the QPC. We find in particular the same universality class as the weak tunneling between fractional quantum Hall edge states at filling $\nu=1/3$ described by  a boundary sine-Gordon model~\cite{Kane_1992, Chamon_1995, Fendley_PRL_1995, Fendley_PRB_1995}. This is readily understood by considering the simplified case where only one QPC is weakly reflecting electrons, the other two QPC being completely open. The two open QPC in series define a dynamical Coulomb blockade environment with impedance $R_s = 2 R_q$ ($R_q = h/e^2$ is the quantum unit of resistance) for the third, a model that has a known mapping~\cite{Safi2004,Pierre_2018a} to the physics of quasiparticle tunneling in a fractional $\nu=1/(1+R_s/R_q) = 1/3$ state, and predict a decreasing conductance as the temperature is lowered. The same model further holds when all three QPC weakly reflect.

With the analytical description of the quantum critical point, we retrieve many features discussed in Ref.~\cite{pouse2021} such as: the residual fractional entropy, the scaling exponents close to the triple points and the shape of the conductance as function of the plunger gate voltages. We also predict the emission of fractional charges~\cite{Kamata2014,berg2009,beri2017,Inoue2014b,Sela_2018} $e^* = e/3$, to be extracted from shot noise measurements~\cite{Saminadayar1997,DePicciotto1997}, despite being in the integer quantum Hall regime~\cite{Feldman_2021,lee2020,morel2021}.

The organization of paper is as follows. In Sec.~\ref{secII}
we discuss the Hamiltonian formulation of two-sites quantum island in the language of bosonization. The effective Hamiltonian at weak backscattering regime is addressed in Sec.~\ref{secIII} in terms of a boundary sine-Gordon model. We outline the calculations for charge current, noise and the evolution of the triple points in Sec.~\ref{secIV}. In Sec.~\ref{secV}, we present details on the Bethe ansatz solution for two-sites quantum island and discuss the comparison of our results with the recent experiment~\cite{pouse2021}. Sec.~\ref{secVI} contains a brief outline on the generalization to the multi-sites cases. We summarize our findings in Sec.~\ref{secVII}. Mathematical details of our calculations are deferred to the Appendices.

\section{Nearly ballistic model}\label{secII}

\begin{figure}
\centering\includegraphics[width=\columnwidth]{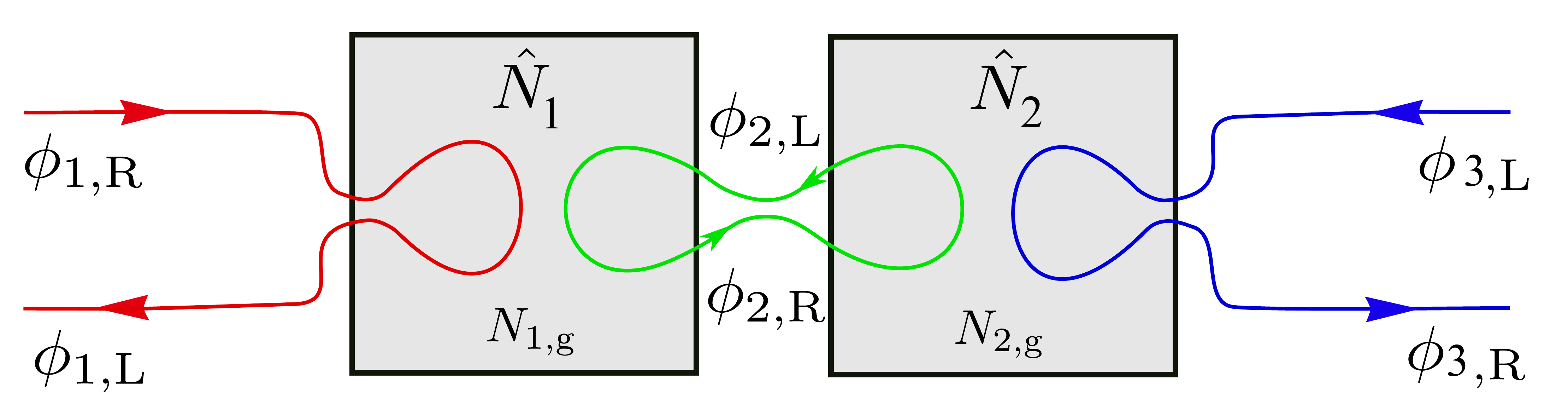}
\caption{Schematic representation of the two-site experiment in Ref.~\cite{pouse2021} A pair of quantum Hall edges connects two floating islands charge-controlled by nearby gate voltages. Two additional external pairs of quantum Hall edge channels contact the island to source and drain. In addition, three QPCs on both sides of each island tune the transmissions of electrons within the different parts.
\label{fig1} }
\end{figure}

We consider the geometry realized in the experiment of Ref.~\cite{pouse2021} and illustrated in Fig.~\ref{fig1}. It comprises three pairs of counter-propagating quantum Hall edges partially covered by two metallic islands. The chiral edge states are best described with bosonization~\cite{Gogolin_book,Giamarchi2003} with the total Hamiltonian $H = H_0 + H_{\rm C} + H_{\rm BS}$. The Hamiltonian $H_0$ governing the propagation of edge states writes
\begin{equation}
H_0 = \frac{v_{\rm F}}{4\pi}\sum_{j=1}^3\int^\infty_{-\infty} dx\Big[\left(\partial_x\phi_{j,{\rm R}}\right)^2+\left(\partial_x\phi_{j,{\rm L}}\right)^2\Big],
\end{equation}
where $\phi_{j,{\rm L/R}}$ represents the bosonic field corresponding to the incoming/outgoing chiral fermions and $v_{\rm F}$ stands for the Fermi-velocity. $H_{\rm c}$ describes the charging  energy of the floating islands
\begin{equation}
H_{\rm C} =  E_{\rm c}\left(\hat{N}_{\rm 1}^2+\hat{N}_{\rm 2}^2\right),
\end{equation}
where for simplicity, the island capacitances $C$ are chosen equal
\footnote{Importantly, the charging energies could be different without affecting any of the universal results we obtain.} 
and $E_{\rm c} = e^2/2 C$. The number densities in the left and right island $\hat{N}_{\rm 1/2}$ are expressed in terms of corresponding gate voltages $N_{\rm 1/2, g}$, controlled by plunger gates, and charge densities $\rho_{j, \alpha}=\frac{1}{2\pi}\partial_x\phi_{j,\alpha}$ such that
\begin{align}
\hat{N}_{\rm 1/2}&=\int^\infty_0 dx\Big[\rho_{\rm 1/2, R}(x)-\rho_{\rm 1/2, L}(x)\Big]\nonumber\\
&+\int^0_{-\infty} dx\Big[\rho_{\rm 2/3, R}(x)-\rho_{\rm 2/3, L}(x)\Big]-N_{\rm 1/2, g}.
\end{align}
The above expressions can be simplified as
\begin{align}\label{nop}
\hat{N}_{\rm 1/2}&=\frac{1}{2\pi}\left(-\delta\phi_{1/2}+\delta\phi_{2/3}-2\pi N_{\rm {1/2}, g}\right),
\end{align}
where we introduce new symbols
\begin{equation}
\delta\phi_j\equiv \phi_{j, {\rm R}}-\phi_{j, {\rm L}}.
\end{equation}
Replacing the finite portions of quantum Hall edges beneath the islands as semi-infinite lines is a standard description first introduced in Refs.~\cite{Matveev1995,Furusaki1995b}. It is justified in the experiments by the very long dwell time of electrons in the metallic islands due to the strong mismatch between high metallic density of states and few outgoing edge channels~\cite{Jezouin2013b}. In contrast, we assume that the two middle chiral edges are fully covered by the left and right islands and neglect the fact that there is some region of space where they are uncovered. It works as long as this uncovered region has a size corresponding to energies well above all other energy scales.

In addition to $H_0$ and $H_{\rm C}$, the Hamiltonian also has a part describing the weak reflection of the three inter-spaced QPCs,
\begin{equation}\label{backscattering}
H_{\rm BS} = \sum_{j=1}^3\frac{D|r_j|}{\pi}\cos\delta\phi_j,
\end{equation}
nonlinear in the bosonic fields, $|r_j|$ are the amplitudes of the corresponding dimensionless reflection coefficients. $D$ is a high-energy scale, or bandwidth, introduced in the bosonization framework. It is necessary at an intermediate step in the formalism but eventually disappears from all practical observable.

\section{Charge average and the sine-Gordon model}\label{secIII}

In the absence of backscattering at the QPC, $|r_j|=0$, the Coulomb blockade induced by the charging energy term $H_{\rm C}$ is entirely suppressed as the flow of electrons becomes continuous and there is no charge granularity. Electrons are continuously entering the island regions before being evenly distributed among the output channels. In the bosonization language, this uninterrupted flow of electrons is described by the quadratic Hamiltonian $H_0 + H_{\rm C}$ with  plasmonic excitations traveling along the chiral edges and scattered at the entrances of the island~\cite{safi1995,safi1999}. It is also represented by an equivalent quantum circuit connecting transmission lines with impedance $R_q$ by capacitors $C$~\cite{morel2021}.

Following Refs.~\cite{Slobodeniuk_2013, Sukhorukov_2016, morel2021}, we use Heisenberg equations of motion to discuss the scattering of plasmonic modes. Their explicit form is detailed in appendix~\ref{appendixA} together with their solution. Solving these equations, we find the following expression for the bosonic modes $\delta \phi_j$,
\begin{equation}\label{scattering}
\begin{pmatrix}
\delta\phi_1\\
\delta\phi_2\\
\delta\phi_3
\end{pmatrix}=\mathbb{M}\begin{pmatrix}
\delta\phi_A^{0}\\
\delta\phi_B^{0}\\
\delta\phi_C^{0}
\end{pmatrix}+\frac{2\pi}{3}\left(
\begin{array}{c}
 -2 N_{\rm 1, g}-N_{\rm 2, g}   \\
 N_{\rm 1, g}-N_{\rm 2, g}   \\
  N_{\rm 1, g}+2 N_{\rm 2, g}  \\
\end{array}
\right).
\end{equation}
The $3 \times 3$ scattering matrix ${\mathbb{M}}$ is given in appendix~\ref{appendixA} and $\tau_c = \pi \hbar/E_c$ is the Heisenberg time associated with the charging energy $E_c$. Eq.~\eqref{scattering} is written in terms of three incoming fields $\delta \phi^0_{A}$, $\delta \phi^0_{B}$,
$\delta \phi^0_{C}$. These three fields are themselves linear combinations of the incoming fields $\delta \phi^0_{1}$, $\delta \phi^0_{2}$,
$\delta \phi^0_{3}$ in the three quantum Hall regions. They have standard commutation relations 
$[ \phi^0_{\alpha, {\rm R/L}} (t), \phi^0_{\alpha, {\rm R/L}} (t')] =-i\pi{\rm sgn}(t-t') $,
with $\alpha=A,B,C$, and, as incoming fields, they originate from a thermalized source with the mean occupancy
\begin{align}
\Big<\delta\phi^0_{\alpha}(\omega) \delta\phi^0_{\alpha}(\omega')\Big> =\frac{4\pi^2}{\omega'}n_{\rm B}\left(\frac{\hbar\omega'}{k_B T}\right)  \delta(\omega+\omega'),
\end{align}
involving the Bose distribution $n_{\rm B}(x)=(e^x-1)^{-1}$ with temperature $T$. 

For energies well below the charging energy $\omega \tau_c \ll 1$, the scattering matrix takes a simple form
\begin{equation}\label{app2}
\lim_{\omega\tau_c\ll 1}\mathbb{M}=-\frac{1}{\sqrt{3}}\begin{pmatrix}
 1 & 0 & 0\\
 1 & 0 & 0\\
 1 & 0 & 0
\end{pmatrix}
\end{equation}
discarding effectively the two fields $\delta \phi^0_{B}$,
$\delta \phi^0_{C}$ in Eq.~\eqref{scattering}. These two fields are thus identified as charge modes that are gapped by the electrostatic charging energy on the two islands. Quite generally, we expect that a series of $N$ capacitive islands would induce a set of $N$ gapped bosonic modes. Hence, we are left with a single gapless mode $\delta \phi^0_{A}$ at low energy that will determine the transport properties throughout the device.

We now include the reflections of the QPCs given by Eq.~\eqref{backscattering}. In the interaction picture, we replace the fields in Eq.~\eqref{backscattering} using Eq.~\eqref{scattering}. At low energy ($<E_{\rm c}$), it is further justified~\cite{Martin_2005} to average $H_{\rm BS} (t)$ over the gapped fields $\delta \phi^0_{B}$ and $\delta \phi^0_{C}$. For instance, the first term writes
\begin{equation}\label{app4}
\cos \delta \phi_1=\frac{1}{2}e^{-\frac{1}{2}\langle \delta\phi_1^2(t)\rangle_{\rm HE}}\!\!\left[e^{i\left[ \frac{\delta\phi^0_A}{\sqrt{3}}+\frac{2\pi}{3}\;(2 N_{\rm 1, g}{+}N_{\rm 2, g})\right]}{+}{\rm h.c.}\right],
\end{equation}
where the average $\langle \rangle_{\rm HE}$ only contains the $\delta \phi^0_{B/C}$ fields. We leave the details of the average calculation to appendix~\ref{appendixB} and quote only the final result,
\begin{equation}\label{app5}
H_{\rm BS}=|r|\;D^{1/3}\;E^{2/3}_{\rm c}\cos\Big(\frac{\delta\phi^0_A}{\sqrt{3}}+\Theta\Big),
\end{equation}
where the effective reflection coefficient $r$ involves a coherent sum over the reflections at the three different QPCs,
\begin{align}\label{rA}
r\equiv  \!\!\frac{\left(3 e^\gamma/\pi\right)^{\frac{2}{3}}}{\pi\sqrt{3}}\!\Bigg(\!&|r_1| e^{\frac{2\pi i}{3}\left(2 \ngo{+}\ngt\right)}
{+}\sqrt{3}|r_2| e^{-\frac{2\pi i}{3}\left( \ngo{-}\ngt\right)}\nonumber\\
&+|r_3| e^{-\frac{2\pi i}{3}\left( \ngo+2\ngt\right)}\Bigg)=|r|e^{i\Theta}.
\end{align}
$\delta \phi_A^0$ is the only chargeless mode, corresponding to a specific plasmonic motion  that is not influenced by the charging energy of the islands. It survives alone for energies below $E_{\rm c}$ and the effective low-energy Hamiltonian that emerges after charge averaging is 
\begin{align}\label{heff}
H_{\rm eff} = \frac{v_{\rm F}}{4\pi} &\int^\infty_{-\infty} dx\Big[\left(\partial_x\phi^0_{A,{\rm R}}\right)^2+\left(\partial_x\phi^0_{A,{\rm L}}\right)^2\Big]\nonumber\\
&+|r|\;D^{1/3}\;E^{2/3}_{\rm c}\cos\Big(\frac{\delta\phi^0_A}{\sqrt{3}}+\Theta\Big),
\end{align}
whereas the current operator at the right output of the device (as detailed in appendix) reads
\begin{equation}\label{currentini}
\hat{I}  \equiv \hat{\mathcal{I}}_3(t)=-\frac{e}{2\pi}\partial_t\delta\phi_3(t)=\frac{e}{2\pi}\frac{1}{\sqrt{3}}\partial_t\delta\phi_A^0(t).
\end{equation}
 Remarkably, the model is a boundary sine-Gordon model, similar to an impurity~\cite{Kane_1992} in a one-dimensional electron liquid with the Tomonaga-Luttinger parameter $K=1/3$ or the weak quasiparticle tunneling between two edges of a fractional quantum Hall state~\cite{Fendley_PRL_1995,Fendley_PRB_1995,Chamon_1995} at filling $\nu=1/3$.

Given a set of QPC transmissions, triple points are defined by specific values of the left and right gate voltages for which the coherent sum in Eq.~\eqref{rA} vanishes, $r=0$. At those points, backscattering processes interfere destructively and a uninterrupted noiseless flow of electron is recovered with maximum conductance. In this case, the effective Hamiltonian~\eqref{heff} is in fact quadratic and readily solvable.

\section{Current, noise and triple points}\label{secIV}

We consider a source-drain geometry where we apply a voltage on the edge channel $1$ and measure the current $\hat{I} \equiv \hat{\mathcal{I}}_3$ in channel $3$. Solving the Heisenberg equations, we obtain that the input field $\delta \phi_1^0 (t)$ is essentially shifted by the number of emitted temporal wavepackets $e V t/\hbar$ during the time interval $t$, which translates into a shift of $e V t/\sqrt{3} \hbar$ for $\delta \phi_A^0 (t)$, and an average current $I = \langle \hat{I} \rangle$
\begin{equation}\label{currentI}
\begin{split}
I & =  \frac{e}{2 \pi \sqrt{3}} \partial_t \left(  \langle \delta \phi_A^0 (t) \rangle + \frac{e V t}{\sqrt{3}  \hbar} \right) \\[2mm]
& = \frac{e^2 V}{3 h} + \frac{e}{2 \pi} \frac{\partial_t \langle  \delta \phi_A^0 (t) \rangle}{\sqrt{3}}.
\end{split}
\end{equation}
At the triple point $r=0$, the Hamiltonian Eq.~\eqref{heff} is quadratic so that $\langle  \delta \phi_A^0 (t) \rangle = 0$, yielding from Eq.~\eqref{currentI} the maximum conductance $G=I/V= G_{\rm max} = e^2/3 h$ as also predicted from NRG calculations~\cite{pouse2021} in the weak tunneling/strong backscattering regime. This conductance can be interpreted  as the series addition of three quantum resistances $R_q$, each QPC being effectively ballistic. The triple point corresponds in fact to the quantum critical point identified in the experiment and NRG calculations of Ref.~\cite{pouse2021}.

Away from the triple point, $r\ne 0$, the non-linear cosine term in Eq.~\eqref{heff} shifts the average bosonic field $\langle  \delta \phi_A^0 (t) \rangle$ to a non-zero value such that conductance becomes non-linear and departs from $e^2/3 h$. This perturbation drives the system away from the quantum critical point and the conductance decreases with decreasing voltage/temperature towards a low-energy fixed point with zero conductance.

\subsection{Fractional charges at weak reflection}

At intermediate energies, the reflection term in Eq.~\eqref{heff} can be treated by perturbation theory. As discussed in appendix, the current expands as $\hat{I} = \hat{I}_0 + \hat{I}_1 + \hat{I}_2$ with
\begin{align}
\hat{I}_0 &=\hat{\mathcal{I}}_3(t) = \frac{e}{2\pi}\frac{1}{\sqrt{3}}\partial_t\delta\phi_A^0(t),\nonumber\\
\hat{I}_1 &=\frac{i}{\hbar}\int^t_{-\infty}dt'\Big[ H_{\rm BS}(t'), \hat{\mathcal{I}}_3(t)\Big],\\
\hat{I}_2 &=-\frac{1}{\hbar^2}\int^t_{-\infty} dt'\int^{t'}_{-\infty}dt''\Big[H_{\rm BS}(t''), \Big[H_{\rm BS}(t'), \hat{\mathcal{I}}_3(t)\Big]\Big]\nonumber.
\end{align}
with $\langle \hat{I}_0 \rangle = e^2 V/3 h$. After taking the quantum average at zero temperature and finite bias voltage, we obtain the leading terms for the current
\begin{equation}\label{current}
I=\frac{1}{3}\frac{e^2 V}{h}\Bigg[1-|r|^2\left(\frac{\overline{E}_{\rm c}}{eV}\right)^{4/3}\frac{\pi^2}{\Gamma\left(2/3\right)}\Bigg],
\end{equation}
with $\overline{E}_{\rm c}=3^{1/4}E_{\rm c}$, and the shot noise
\begin{equation}
S=\frac{e}{9}\frac{e^2V}{h}|r|^2\left(\frac{\overline{E}_{\rm c}}{eV}\right)^{4/3}\frac{\pi^2}{\Gamma\left(2/3\right)}.
\end{equation}
The Fano factor for the reflected current is then 
\begin{equation}
F =\frac{S}{I(|r|=0)-I}=\frac{1}{3},
 \end{equation}
corresponding to the backscattering of fractional charges $e^*=e/3$. Eq.~\eqref{current} can be written in a more suggestive form
\begin{equation}\label{current2}
I=\frac{1}{3}\frac{e^2 V}{h}\Bigg[1- a \left(\frac{k_B T_*}{eV}\right)^{4/3}\Bigg],
\end{equation}
with the coefficient $a=(\sqrt{\pi}/6) (\Gamma\left(\frac{1}{3}\right)/ \Gamma\left(\frac{5}{6}\right))$, by introducing the temperature scale~\footnote{the choice of the numerical factors $a$ and $b$ is somewhat arbitrary and only their combination $a b^{4/3}$ is fixed by Eq.~\eqref{current}. Our choice is guided by having a compact expression for the differential conductance in Sec.~\ref{sec-BA}.}
\begin{equation}\label{tstar}
T_*=b \frac{|r|^{3/2} \overline{E}_{\rm c}}{k_B}
\end{equation}
with $b = \left[3\sqrt{3\pi}\Gamma\left(5/6\right)\right]^{3/4}$. Eq.~\eqref{current2}, valid in the perturbative regime $e V \gg k_B T_*$, gives the onset of the crossover to the low-voltage regime. While $T_*$ itself is a non-universal scale depending on microscopic details ({\it e.g.} high-energy details of the electronic dispersion relation, irrelevant processes,...), the current is found to be a universal function of $eV/k_BT_*$ and $T/T_*$. The full crossover  function of $I(e V/k_B T_*,T/T_*)$  is determined from the Bethe ansatz solution of the Sine-Gordon model defined by Eq.~\eqref{heff}, as later discussed in Sec.~\ref{sec-BA}.

\subsection{Triple points}
\begin{figure}
\begin{center}
\includegraphics[scale=0.45]{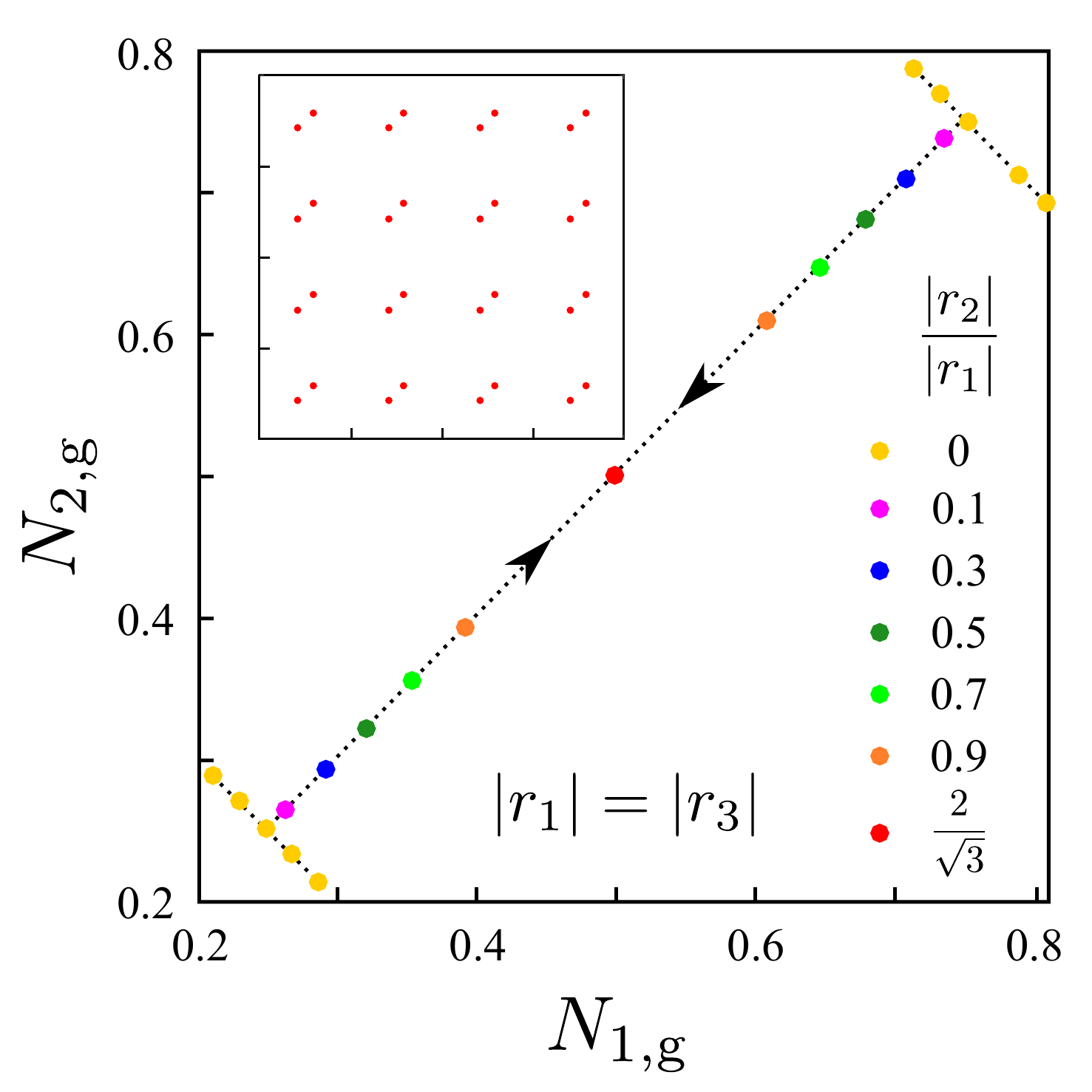}
\caption{Triple points $r=0$ represented in the plane of gate voltages for increasing $|r_2|/|r_1|$ at $|r_1|=|r_3|$. A single unit cell is represented (the inset shows the periodic pattern for $|r_2|=|r_1|$). The two triple points originate from $\ngo=(1-2\ngt)/2$ at $|r_2|/|r_1|=0$ and move towards each other as $|r_2|/|r_1|$ is increased until they annihilate at $|r_2|/|r_1| =2/\sqrt3 $. Triple points no longer exists above this critical value.
\label{fig2} }
\end{center}
\end{figure}

\begin{figure}
\begin{center}
\includegraphics[scale=0.45]{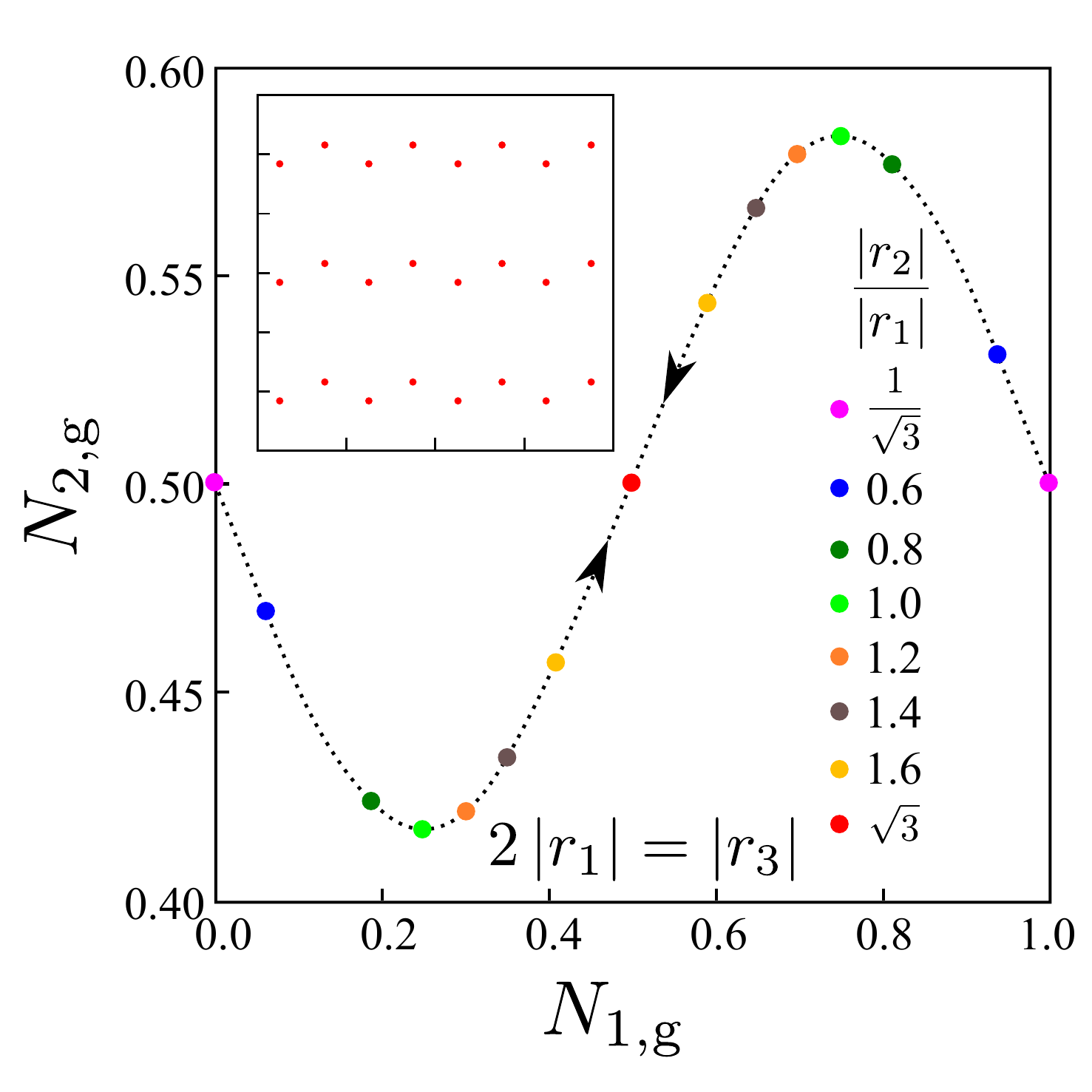}
\caption{Triple points $r/|r_1|=0$ movement in the plane of gate voltages for increasing $|r_2|/|r_1|$ for the asymmetrical configuration with $2|r_1|=|r_3|$.The inset shows the periodic pattern with enlarged gate voltages for $|r_2|=|r_1|$.
\label{fig3}}
\end{center}
\end{figure}

The temperature $T_*$ from Eq.~\eqref{tstar} sets the energy scale for the crossover between the high and low temperature or bias voltage limits. It is governed by the effective reflection coefficient $r$ and therefore vanishes at the triple points where the conductance is $e^2/3 h$, independent of the temperature or bias voltage. Fig.~\ref{fig2} maps out the positions of the triple points for the symmetric configuration $|r_1|= |r_3|$ as $|r_2|/|r_1|$ evolves. The triple points form a periodic lattice with two sites per unit cell when the left and right gate voltages are varied. They only exist for $|r_2|/|r_1|<2\sqrt3$. For  $|r_2|/|r_1|>2\sqrt3$, the triple points disappear: the electron-mediated coupling between the two islands becomes too weak and the two islands couple preferentially to the source and drain leads which suppresses transport at low energy.  
In the asymmetric case $|r_1|\neq |r_3|$, triple points only exist within upper and lower threshold of $|r_2|/|r_1|$ determined by the ratio $|r_3|/|r_1|$,
\begin{align}
\left.\frac{|r_2|}{|r_1|}\right|_{\rm lower} &=\begin{cases}
\frac{ \frac{|r_3|}{|r_1|}-1}{\sqrt{3}}, &  \frac{|r_3|}{|r_1|}\geq 1,\\
\frac{ 1-\frac{|r_3|}{|r_1|}}{\sqrt{3}},& \frac{|r_3|}{|r_1|}< 1,
\end{cases}\\
\left.\frac{|r_2|}{|r_1|}\right|_{\rm upper} &=\frac{ \frac{|r_3|}{|r_1|}+1}{\sqrt{3}}.
\end{align}
Beyond these thresholds, pair of triple points eventually meet and disappear. The positions of triple points in the plane defined by gate voltages is depicted in Fig.~\ref{fig3}.
From Fig~\ref{fig2}, it is seen that the evolution of triple point with varying $|r_2|/|r_1|$ for the symmetrical configuration is entirely on the diagonal plane $\left(\ngo,\ngt\right)=\left(N_{\rm g}, N_{\rm g}\right)$ such that
\begin{equation}
N_{\rm 1, g}=\frac{1}{2\pi}\arccos\left[-\frac{\sqrt{3}\; \frac{|r_2|}{|r_1|}}{2}\right]=N_{\rm 2, g}=N_{\rm g}.
\end{equation}
Then the expansion of Eq.~\eqref{rA} around the triple points in the symmetric case writes
\begin{align}\label{ellipse}
\left|\frac{r}{|r_1|}\right|^2 =4\pi^2&\Bigg(\frac{\left(3 e^\gamma/\pi\right)^{\frac{2}{3}}}{\pi\sqrt{3}}\Bigg)^2\Bigg((\delta N_1)^2+ (\delta N_2)^2\nonumber\\
&+\left[2-3  \left(\frac{|r_2|}{|r_1|}\right)^2 \right]\delta N_1\;\delta N_2\Bigg),
\end{align}
where we define two small parameters $\delta N_{1/2} =N_{\rm g, 1/2}-N_{\rm g}$. The Eq.~\eqref{ellipse} draws an ellipse squashed along the $\ngo = \ngt$ direction, or a circle for the special case $|r_2|/|r_1| =\sqrt{2/3} $. The two triple points are in fact symmetrically positioned with respect to the central configuration of gate voltages $\ngo = \ngt=1/2$. Along the line $\ngo = \ngt$, $r$ thus develops a double well form with zeros at the triple points and a local maximum at the center $\ngo = \ngt=1/2$.

\section{Crossover function}\label{secV}
\subsection{Bethe ansatz solution}\label{sec-BA}

The boundary sine-Gordon model of Eq.~\eqref{heff}, together with the definition of the current observable Eq.~\eqref{currentini} and the coupling to the bias voltage described by the time-dependent shift in the boson of Eq.(\ref{currentI}) defines the canonical out-of-equilibrium boundary sine-Gordon model. This model also faithfully describes the 
 the tunneling between $\nu=1/3$ fractional quantum Hall states, or between L\"uttinger liquids with parameter $K=\frac13$, and in this
 context it has been shown that it allows for an exact out-of-equilibrium solution using Bethe Ansatz~\cite{Fendley_PRL_1995,Fendley_PRB_1995}. The main features of the Bethe Ansatz solution are summarized in Appendix~\ref{appendixC}. It is characterized by a single temperature (energy) scale $T_*$ entering all universal functions for physical observables such as the current, the noise, etc. Quite generally, $T_*$ depends on a combination of the sine-Gordon coupling constant $r$ and the ultraviolet cutoff of the model. In our case, it is readily obtained by matching the Bethe Ansatz high-voltage $e V \gg k_B T_*$ expansion of the current with its microscopic perturbative evaluation  in Eq.~\eqref{current}. The resulting expression for the temperature scale $T_*$ has been anticipated and given in Eq.~\eqref{tstar}. As expected, the different powers of $D$ cancel out and the intermediate cutoff energy $D$ finally drops out from the expression of $T_*$, and thus of all observables.

It results that the non-linear differential conductance exhibits the scaling form
\begin{equation}\label{diffV}
G (V,T) = \frac{1}{3}\frac{e^2}{ h} g \left( \frac{e V}{k_B T_*} , \frac{T}{T_*} \right),
\end{equation}
with $g(x,y)$ a universal function.
At zero temperature, an analytical expression for $g$ is derived from the Wiener-Hopf technique~\cite{Fendley_PRB_1995}
\begin{equation}
g(x,0)=\begin{cases}
1-\sum\limits _{n=1}^{\infty } \frac{(-1)^{n+1}\sqrt{\pi }  \Gamma \left(\frac{n}{3}\right)}{3\Gamma (n) \Gamma \left(\frac{1}{2}-\frac{2 n}{3}\right)} x^{-\frac{4 n}{3}},&x>\left(\frac{4}{27}\right)^{1/4}\\
\sum\limits _{n=1}^{\infty } \frac{(-1)^{n+1}3 \sqrt{\pi }  \Gamma (3 n)}{\Gamma (n) \Gamma \left(\frac{1}{2}+2 n\right)} x^{4 n}, &x<\left(\frac{4}{27}\right)^{1/4},
\end{cases}
\end{equation}
and describes the universal crossover from low to high voltage.
As already noticed~\cite{Fendley_PRB_1995,Kane_1992,Lesage1999,Fendley1998,duprez2021}, it exhibits a duality between its low- and high-energy expansions. Interestingly, the differential conductance 
is super-ballistic at large voltage as it exceeds~\cite{Fendley_PRB_1995} the ballistic limit $e^2/(3 h)$, before falling off down to zero at vanishing bias. The corresponding trace is shown in Fig.\ref{ucurvev} in the appendix~\ref{appendixC}.

\begin{figure}
\begin{center}
\includegraphics[scale=0.4]{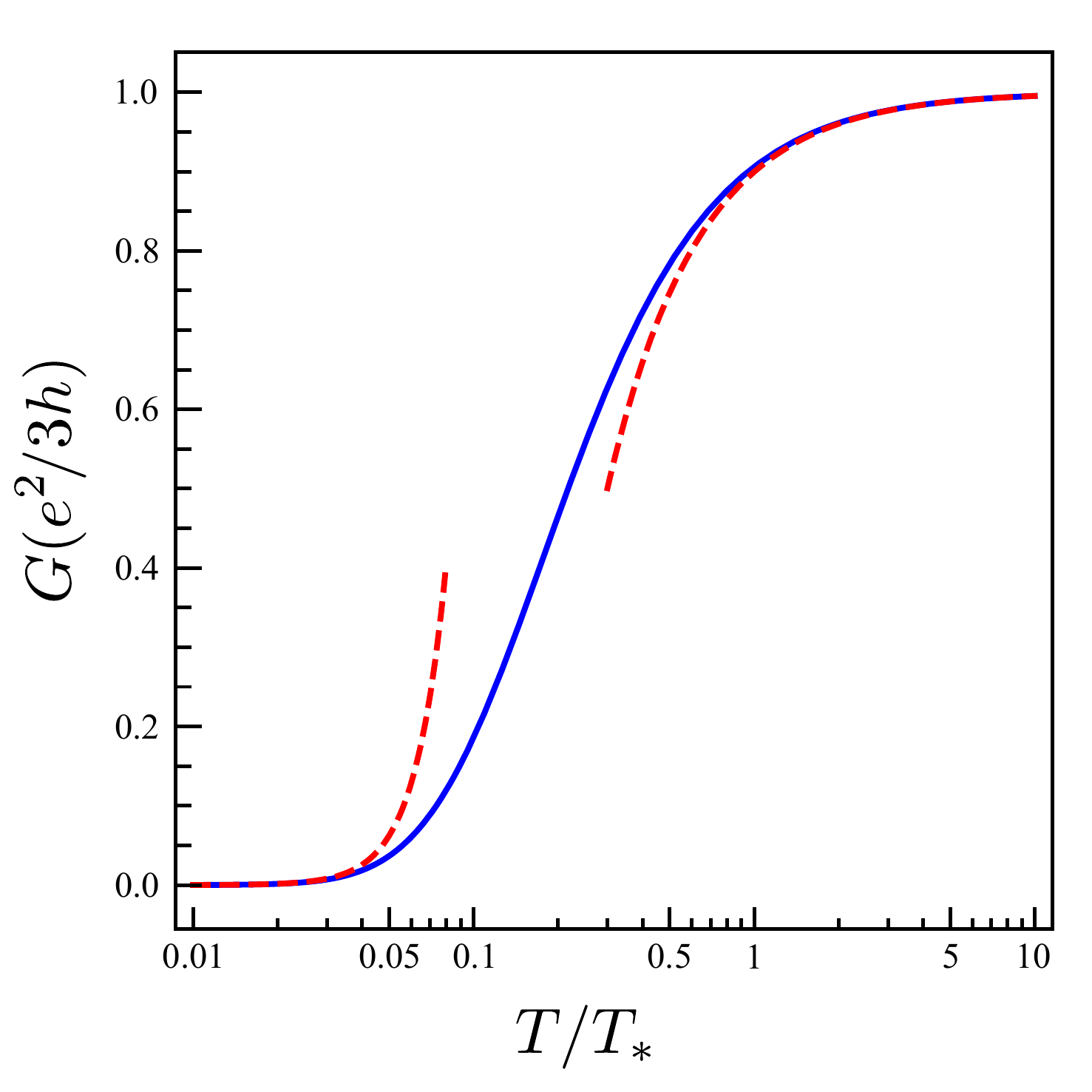}
\caption{Universal linear conductance  in units of $e^2/3 h$ as function of the temperature ratio $T/T_*$. The conductance (blue plain line) is obtained from the thermodynamic Bethe ansatz (see appendix~\ref{appendixC}) and compared to its low- and high-temperature asymptotes (red dashed lines) given in Eq.~\eqref{asymptotesG}.}\label{ucurvet}
\end{center}
\end{figure}
The thermodynamic Bethe ansatz~\cite{Fendley_PRL_1995,Fendley_PRB_1995} provides the solution at finite temperature and is detailed in appendix~\ref{appendixC}.
Observables are then given by universal functions of the temperature ratio $T/T_*$. The resulting linear conductance is shown in Fig.~\ref{ucurvet}. Its low- and high-temperature  asymptotic behaviors were derived in Ref.~\cite{Boulat2019}:
\begin{equation}
\begin{split}
g(0,y\ll 1) &\simeq  \frac{32}{5}\; ( 2\pi y)^4, \\[2mm]  
g(0,y\gg 1) &\simeq 1-
\frac{\sqrt\pi \Gamma(\frac13)^3}{6\Gamma(\frac56)}
\; \frac{1}{(6\pi y)^{4/3}}.
\label{asymptotesG}
\end{split}
\end{equation}
The universal function $g(0,y)$ describes a monotonous crossover between the high-temperature unitary conductance $e^2/(3 h)$ and the vanishing zero-temperature conductance. The $T^4$ scaling at low temperature in Eq.~\eqref{asymptotesG} can be readily understood from an inelastic co-tunneling perspective~\cite{averin1989,averin1990,Averin1992,matveev1996}. The transfer of one electron from the left to the right leads occurs via virtual processes where each island is excited with an electron-hole pair. The phase space for each pair is proportional to $T^2$ yielding an overall $T^4$ scaling for the conductance. In the bosonization language, the operator responsible for the $T^4$ scaling is $e^{i \sqrt{3} \delta \theta_A}$, the dual to the cosine backscattering term of Eq.~\eqref{app5}, where $\delta \theta_A$ is the canonical conjugate to the field $\delta \phi_A$. It can also be written as
\begin{equation}
e^{i \sqrt{3} \delta \theta_A} = e^{i \delta \theta_1}      e^{i \delta \theta_2} e^{i \delta \theta_3}.
\end{equation}
The vertex operator $e^{i \delta \theta_1}$ corresponds to the transfer of one electron across the first (left) QPC and can be fermionized back as $\psi_{R,1}^\dagger  \psi_{L,1}$. The field $\psi_{L/R,1}$ annihilates an electron on the left (resp. right) side of the first QPC. Using similar refermionization at each QPC, one obtains the leading irrelevant low-energy operator 
\begin{equation}\label{lio}
e^{i \sqrt{3} \delta \theta_A} \propto \psi_{R,1}^\dagger  \psi_{L,1} \psi_{R,2}^\dagger  \psi_{L,2}\psi_{R,3}^\dagger  \psi_{L,3}
\end{equation}
corresponding indeed to the transfer of one electron across the two-island structure together electron-hole excitations in both islands. 

The operator of Eq.~\eqref{lio} involves the fermionic reservoirs around the three QPC symmetrically. This symmetry  originates in fact from the symmetric expression of $\delta \phi_A$ in terms of $\delta \phi_{1,2,3}$ which is preserved along the integrable crossover, yielding eventually the six operators in Eq.~\eqref{lio} and the $T^4$ scaling. Such a symmetry is a physical one in the case of tunneling between two fractional quantum Hall states at $\nu=1/3$ as only electrons, reformed as triplet of quasiparticles, are physically allowed to tunnel at low energy. For our system, this threefold symmetry is not generally preserved. The Bethe-ansatz solution predicts a Fermi liquid state at low temperature but the obtained scaling $T^4$ is not symmetry-protected and indeed a $T^2$ scaling was identified in the NRG calculation of Ref.~\cite{pouse2021}. In contrast, the high-temperature region of Fig.~\ref{ucurvet} is robust and in agreement with the measurements and NRG calculations of Ref.~\cite{pouse2021}.


\subsection{Comparison with Ref.~\cite{pouse2021}}

The geometry of two connected quantum islands discussed in this paper has been realized in the experiment of Ref.~\cite{pouse2021} together with NRG calculations adapted to the limit of weakly transparent QPC - the opposite limit is the focus of our work. 
Nevertheless, as already mentioned in the introduction, the properties of the quantum critical point are expected to be unique, independent of the bare QPC transmissions, and our work can shed light on many findings of Ref.~\cite{pouse2021}.

The main difference between our regime of high QPC transparency and Ref.~\cite{pouse2021} is the absence of a Kondo scaling region. For a weak QPC transparency, a Kondo resonance forms progressively with decreasing temperature, yielding a conductance that 
increases when temperature is lowered and eventually saturates at a value $G_0=\frac{e^2}{3h}$ at the triple point.
The resonance is characterized by the Kondo temperature scale $T_K$. $T_K$ increases with the QPC transparency and the Kondo effect thus disappears when the transparency exceeds a threshold for which $T_K \sim E_c$. In the quasi-ballistic regime, {\it i.e.} almost transparent, many charge states are occupied and there is no Kondo effect. In summary, the conductance at large transparency, shown in Fig.\ref{ucurvet}, is a monotonous and increasing function of temperature, with the characteristic temperature scale $T^*$ whereas, in the weakly transparent case of  Ref.~\cite{pouse2021}, the conductance also starts as an increasing function of $T/T^*$ but then crosses over to a decreasing function of $T/T_K$.

Interestingly, the nature of the relevant perturbation in the vicinity of the quantum critical (triple) point does not depend on the QPC bare transparency. We expect that the $\sim (T/T_*)^{-4/3}$ correction of Eq.~\eqref{asymptotesG} is valid regardless of the transparency and applies as well to the experiment of  Ref.~\cite{pouse2021} under the condition that $T,T^* \ll T_K, E_c$.
A similar discussion with two temperature scales can be found in Ref.~\cite{Sela_2011,Mitchell_2012a,Sela_2016} for the two-channel Kondo model slightly away from charge degeneracy~\cite{Pierre_2015}. 
The $T^{-4/3}$ scaling is in fact clearly observed in the NRG data of Ref.~\cite{pouse2021}, and our Fig.~\ref{ucurvet} coincides well with NRG from high to moderate values of $T/T_*$.

All in all, our approach recovers other findings from Ref.~\cite{pouse2021}. It identifies the nature of the quantum critical point with its leading relevant perturbation: a boundary sine-Gordon model with the Luttinger parameter $K=1/3$. The critical point is characterized~\cite{fendley1994} by the fractional residual entropy $\Delta S = \ln (\sqrt{3} )$, also observed in the NRG calculation of Ref.~\cite{pouse2021}. Our approach also explains the value of the critical exponents in the vicinity of the quantum critical point: from Eqs.~\eqref{tstar} and~\eqref{ellipse}, $T_*$ varies as $\sim \delta N_g^{3/2}$ and $\sim |r_{2}|^{3/2}$, in precise agreement with Ref.~\cite{pouse2021}.

\begin{figure}
\begin{center}
\includegraphics[scale=0.35]{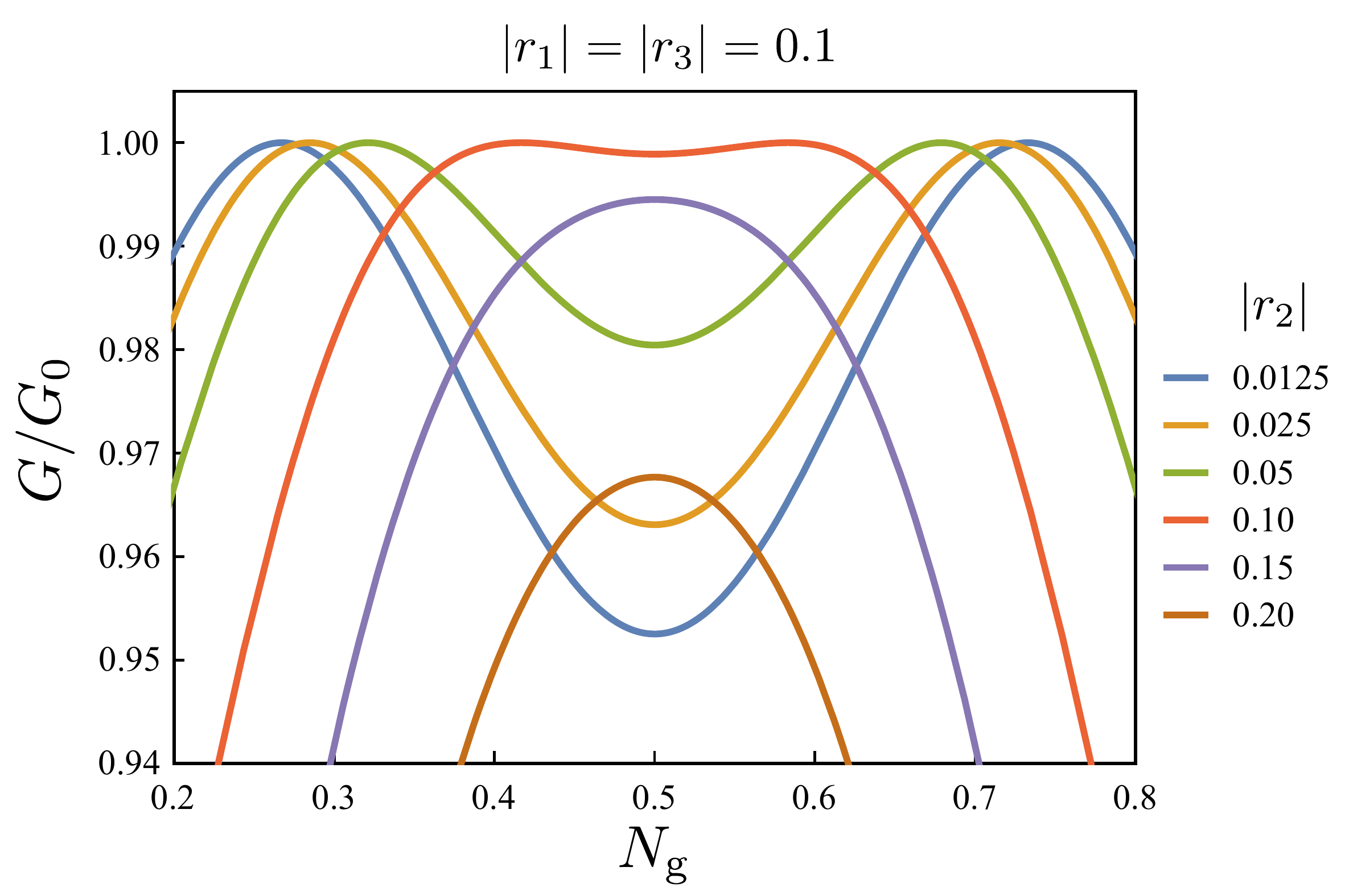}
\caption{The finite temperature differential conductance (in the unit of $G_0=e^2/3h$) as a function of gate voltage $\ngo=\ngt=N_{\rm g}$ for $k_BT/\overline{E}_{\rm c}=0.1$ and reflection coefficients as specified in the plot.}\label{Fig5}
\end{center}
\end{figure}

Since $T_*$ depends explicitly on the left and right gate voltages through Eq.~\eqref{tstar} and Eq.~\eqref{rA}, we can evaluate the linear conductance as a function of the gate voltage $\ngo$ along the symmetric line $\ngo = \ngt$ for a fixed temperature $T$. The result is shown in Figs.~\ref{Fig5} for different values of $|r_2|$, illustrating the competition between the cooperative transport through the two islands via their edge-state mediated coupling on one hand and the current suppression of each individual island due to the relevant QPC backscattering on the other hand. The very same behaviour for the conductance has been measured and calculated via NRG in Ref.~\cite{pouse2021} (see their Figure 2) in the weakly transparent regime, and interpreted as a competition between Kondo screening of each individual island and the antiferromagnetic inter-island binding.
\begin{figure}
\begin{center}
\includegraphics[width=\columnwidth]{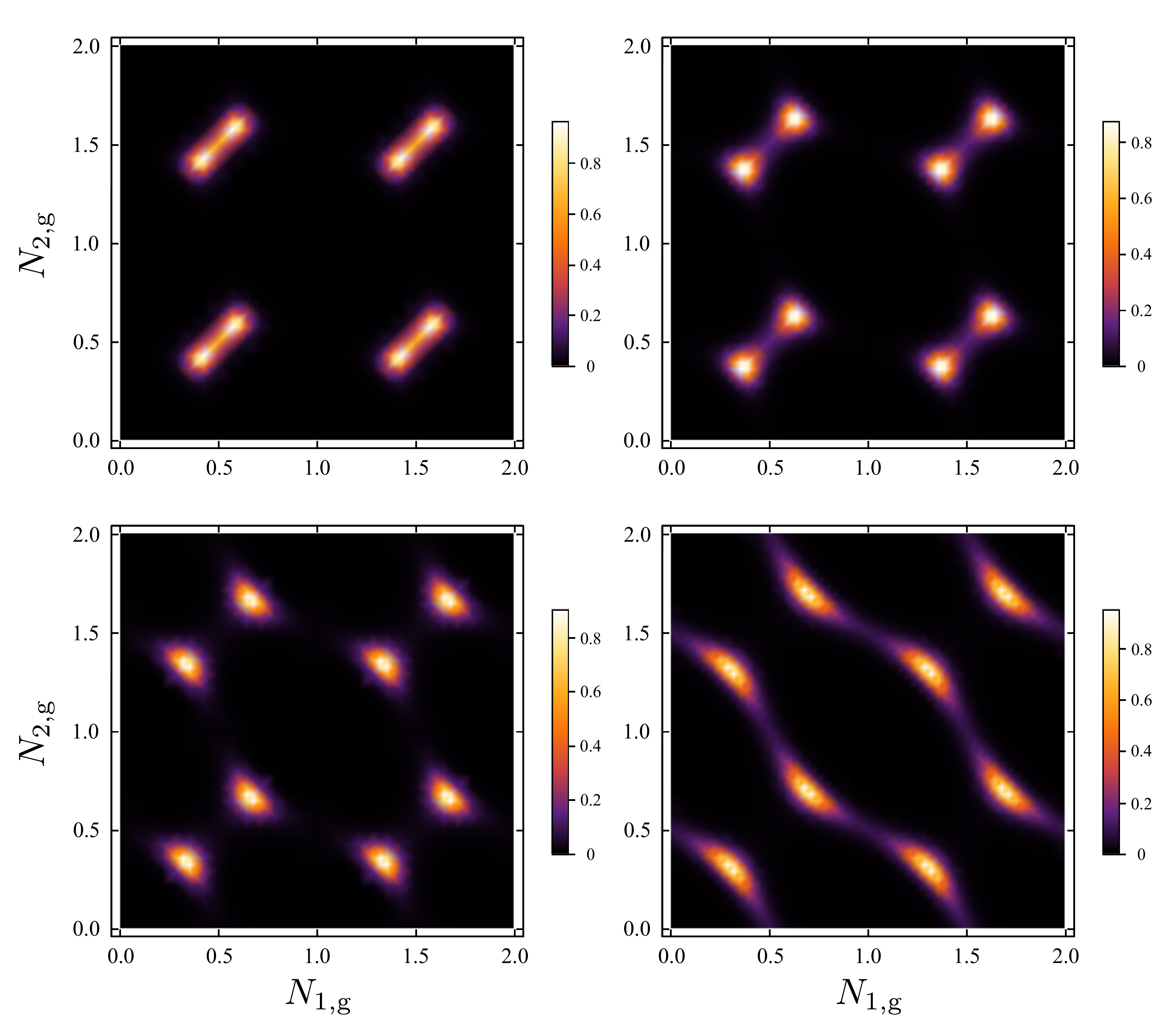}
\caption{Evolution of the differential conductance (in the unit of $e^2/3h$) in the plane defined by the gate voltages $N_{\rm g, 1/2}$ at temperature $k_B T/\overline{E}_{\rm c}=0.001$. The plots from top to bottom correspond to the symmetrical configuration $|r_1|=|r_3|=0.1$ with $|r_2|=0.1, 0.08, 0.06, 0.04$ respectively.}\label{densityplot}
\end{center}
\end{figure}
\section{Generalization to many islands}\label{secVI}
\begin{figure}
\begin{center}
\includegraphics[width=\columnwidth]{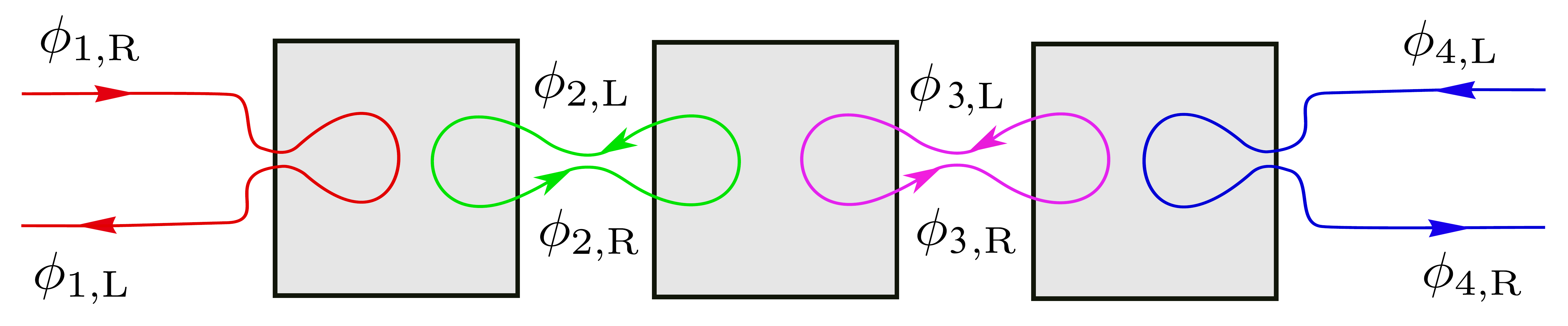}
\caption{Three-sites quantum islands.}\label{3QD}
\end{center}
\end{figure}

The results presented above can be straightforwardly extended to more complex setups such as many quantum islands connected in series by single channel QPCs. We consider as an example the illustrative case of three islands depicted in Fig~\ref{3QD}.
In this case, we express the four bosonic fields $\delta\phi_j$ in terms of the four incoming fields $\delta\phi^0_\alpha$ via a $4\times 4$ scattering matrix $\mathbb{M}$. In the low-energy limit, $\mathbb{M}$ writes
\begin{equation}
\lim_{\omega\tau_c\ll 1}\mathbb{M}=\frac{1}{2}\left(
\begin{array}{cccc}
 1 & 0 & 0 & 0 \\
 1 & 0 & 0 & 0 \\
 1 & 0 & 0 & 0 \\
 1 & 0 & 0 & 0 \\
\end{array}
\right).
\end{equation}
This shows that, out of the four incoming fields, only one $\delta\phi^0_A$ is a gapless mode and the remaining three modes can be integrated out to obtain an effective Hamiltonian energies well below the charging energy, similarly to the previous case of two connected islands. The result of the averaging in the absence of gate voltages gives
\begin{equation}
H'_{\rm BS}=r' D^{1/4} E^{3/4}_{\rm c}\cos\left(\frac{\delta\phi^0_A}{2}\right),
\end{equation}
where the effective reflection coefficient writes
\begin{align}
r'=\frac{\sqrt2}{\pi}&\left(\frac{e^\gamma}{\pi}\right)^{3/4}\Big[\left(|r_1|+|r_4|\right)\left(17-2\sqrt2\right)^{\frac{1}{8\sqrt2}}\nonumber\\
&+\left(17-2\sqrt2\right)^{-\frac{1}{8\sqrt2}}\left(|r_2|+|r_3|\right)\Big].
\end{align}
We find again a boundary sine-Gordon model where the dimension of the relevant operator is $1/4$. The subsequent analysis takes the same form as for two islands and the transport properties are calculated analogously. Computing the current and the noise perturbatively for $T \gg T^*$, where $T^* \sim |r'|^{8/5} E_c/k_B$ is the temperature scale generated by the boundary term, we find the backscattering of fractional charges $e^* = e/4$.

We can readily generalize our setup to $N_{\rm QI}$ islands connected by interspaced individual QPCs. In that case, the theory follows the same line, one obtains a single gapless bosonic mode that survives at energies well below the charging energy. The corresponding scattering matrix $\mathbb{M}$ of size $\left(N_{\rm QI}+1\right)\times\left(N_{\rm QI}+1\right)$ simplifies at low energy with a single non-vanishing column with entries $1/\sqrt{N_{\rm QI}}$. The resulting model for the quantum critical point and its leading relevant perturbation is still a boundary sine-Gordon model with the operator dimension $\nu=1/\left(N_{\rm QI}+1\right)$. Perturbative evaluations of the current and noise give the backscattered fractional charge $e^* = e/\left(N_{\rm QI}+1\right)$.


\section{Conclusion}\label{secVII}
In light of a recent experiment~\cite{pouse2021}, we investigate theoretically the quantum criticality associated with two floating islands connected to three edge channels via quantum point contacts in the integer quantum Hall regime. Assuming high transparency of the quantum point contacts, we unveil the nature of the quantum critical point by establishing an explicit mapping to the boundary sine-Gordon model with Luttinger parameter $K=1/3$. 
From this mapping to the boundary sine-Gordon model, we study the features of different observables reported in the experiment~\cite{pouse2021}. 
We compute the residual fractional entropy, and we also find that the critical exponents for the conductance close to the triple points and the shape of the conductance as a function of the plunger gate voltages are in full agreement with experimental findings.

Our analytical description also demonstrates the competition between Kondo-like screening of each individual island and the cooperative transfer of electrons between them via the conductance measurements. In addition, we report the emission of fractional charges $e^*=e/3$ whose value can be extracted from shot noise
measurements. Our work also sheds light towards extending the recent experiment to charge-Kondo clusters. For multi ($N_{\rm QI}$)-sites quantum islands, we show that the mapping to the boundary sine-Gordon model still holds -- with Luttinger parameter $K=1/\left(N_{\rm QI}+1\right)$ -- and find a fractional charge emission $e^*=e/\left(N_{\rm QI}+1\right)$. 

\section*{Acknowledgment}

We are grateful to F. Pierre, P. Sriram and A. K. Mitchell for inspiring discussions.  This work was supported by the French National Research Agency (project SIMCIRCUIT, ANR-18-CE47-0014-01).

\appendix
\section{Equations of motion for bosonic fields}\label{appendixA}
With number density operators $\hat{N}_{1/2}$ presented in Eq.~\eqref{nop}, we arrive at the equation of motions for the bosonic fields (for details see Ref.~\cite{morel2021})
\begin{align}
\delta\phi_1 &=\delta\phi_1^0-\frac{1}{i\omega\tc}\left(-\delta\phi_1+\delta\phi_2-2\pi N_{\rm 1, g}\right),\nonumber\\
\delta\phi_2 &=\delta\phi_2^0+\frac{1}{i\omega\tc}\left[-\delta\phi_1+2\delta\phi_2-\delta\phi_3+2\pi\left(N_{\rm 2, g}-N_{\rm 1, g}\right)\right],\nonumber\\
\delta\phi_3 &=\delta\phi_3^0+\frac{1}{i\omega\tc}\left(-\delta\phi_2+\delta\phi_3-2\pi N_{\rm 2, g}\right).\label{app1}
\end{align}
By performing orthogonal transformation
\begin{align}
&\begin{pmatrix}
\delta\phi_A^{0}\\
\delta\phi_B^{0}\\
\delta\phi_C^{0}
\end{pmatrix}=
\begin{pmatrix}
-\frac{1}{\sqrt{3}} & \frac{-1}{\sqrt{3}} & \frac{-1}{\sqrt{3}}\\
-\frac{1}{\sqrt{2}} & \frac{1}{\sqrt{2}} & 0\\
-\frac{1}{\sqrt{6}} & \frac{-1}{\sqrt{6}} & \frac{2}{\sqrt{6}}
\end{pmatrix}
\begin{pmatrix}
\delta\phi_1^{0}\\
\delta\phi_2^{0}\\
\delta\phi_3^{0}
\end{pmatrix},
\end{align}
we express the Eq.~\eqref{app1} into the form already presented in Eq.~\eqref{scattering} with the scattering matrix
\begin{equation}\label{app3}
\mathbb{M}= \left(
\begin{array}{ccc}
 -\frac{1}{\sqrt{3}} & -\frac{\omega  \tau _c \left(2 i+\omega  \tau _c\right)}{\sqrt{2} \left(\omega ^2 \tau _c^2+4 i \omega  \tau _c-3\right)} & -\frac{\omega  \tau _c \left(4 i+\omega  \tau _c\right)}{\sqrt{6} \left(\omega ^2 \tau _c^2+4 i \omega  \tau _c-3\right)} \\
 -\frac{1}{\sqrt{3}} & \frac{\omega  \tau _c}{\sqrt{2} \left(3 i+\omega  \tau _c\right)} & -\frac{\omega  \tau _c}{\sqrt{6} \left(3 i+\omega  \tau _c\right)} \\
 -\frac{1}{\sqrt{3}} & \frac{i \omega  \tau _c}{\sqrt{2} \left(\omega ^2 \tau _c^2+4 i \omega  \tau _c-3\right)} & \frac{\omega  \tau _c \left(5 i+2 \omega  \tau _c\right)}{\sqrt{6} \left(\omega ^2 \tau _c^2+4 i \omega  \tau _c-3\right)}
\end{array}
\right),
\end{equation}
whose low energy limit $\omega\tau_c\ll 1$ is given in Eq.~\eqref{app2}. 
\section{Effective Hamiltonian}\label{appendixB}
We start form the Hamiltonian Eq~\eqref{backscattering} accounting for weak backscattering in three QPCs.  From appendix.~\ref{appendixA}, we see that $\delta\phi_i$ entering into Eq~\eqref{backscattering} are expressed in terms of three incoming fields $\delta\phi^0_\alpha$ ($\alpha=A, B,C$). While $\alpha=A$ is the gapless mode, $\alpha=B, C$ represent gapped modes. We then integrated out the gapped modes to arrive at the backscattering Hamiltonian expressed in terms of single gapless mode $\delta\phi^0_A$. In the following, we provide a quick summary on the integrating out of the high-energy modes. From the scattering matrix presented in appendix~\ref{appendixA}, we have for the field $\delta\phi_1$
\begin{align}
\delta\phi_1 = {-}\frac{\delta\phi_A^{0}}{\sqrt{3}}&{-}\frac{2\pi}{3}(2 N_{\rm 1, g}{+}N_{\rm 2, g})-\frac{\omega  \tau _c \left(2 i{+}\omega  \tau _c\right)\;\delta\phi_B^{0}}{\sqrt{2} \left(\omega ^2 \tau _c^2{+}4 i \omega  \tau _c{-}3\right)}\nonumber\\
&-\frac{\omega  \tau _c \left(4 i+\omega  \tau _c\right)\;\delta\phi_C^{0}}{\sqrt{6} \left(\omega ^2 \tau _c^2+4 i \omega  \tau _c-3\right)}.
\end{align}
Above equation results in the cosine term in the backscattering Hamiltonian of the QPC1 as given in Eq.\eqref{app4}. The average over the high energy modes $\langle \delta\phi_1^2(t)\rangle_{\rm HE}$ is obtain by using the input/output scattering formalism developed in Refs.~\cite{Slobodeniuk_2013, Sukhorukov_2016, morel2021} followed by the application of identity
\begin{equation}\nonumber
\Big<\delta\phi^0_{\alpha}(\omega)\left(\delta\phi^0\right)^\dagger_{\alpha'}(\omega')\Big>=2\left[1{+}n_{\rm B}\left(\frac{\hbar\omega}{k_B T}\right)\right]\delta_{\alpha\alpha'}\delta(\omega-\omega').
\end{equation}
At zero temperature, we arrive at the result
\begin{align}
\langle \delta\phi_1^2(t)\rangle_{\rm HE} =\frac{4}{3}\int^\infty_{0}&d\omega\frac{\omega\tau^2_c\left[7+(\omega\tau_c)^2\right]}{\left[1+(\omega\tau_c)^2\right]\left[9+(\omega\tau_c)^2\right]} e^{-\frac{\omega\hbar}{D}}\nonumber\\
&=-\frac{4}{3}\log\left[3^{1/4} e^\gamma \frac{\hbar}{D\tc}\right].
\end{align}
Repeating the same procedure for QPC2 and QPC3, the Eq.~\eqref{backscattering} finally writes into the form of Eq.~\eqref{app5} presented in the main text.
\section{Bethe ansatz solution}\label{appendixC}
\begin{figure}
\begin{center}
\includegraphics[scale=0.4]{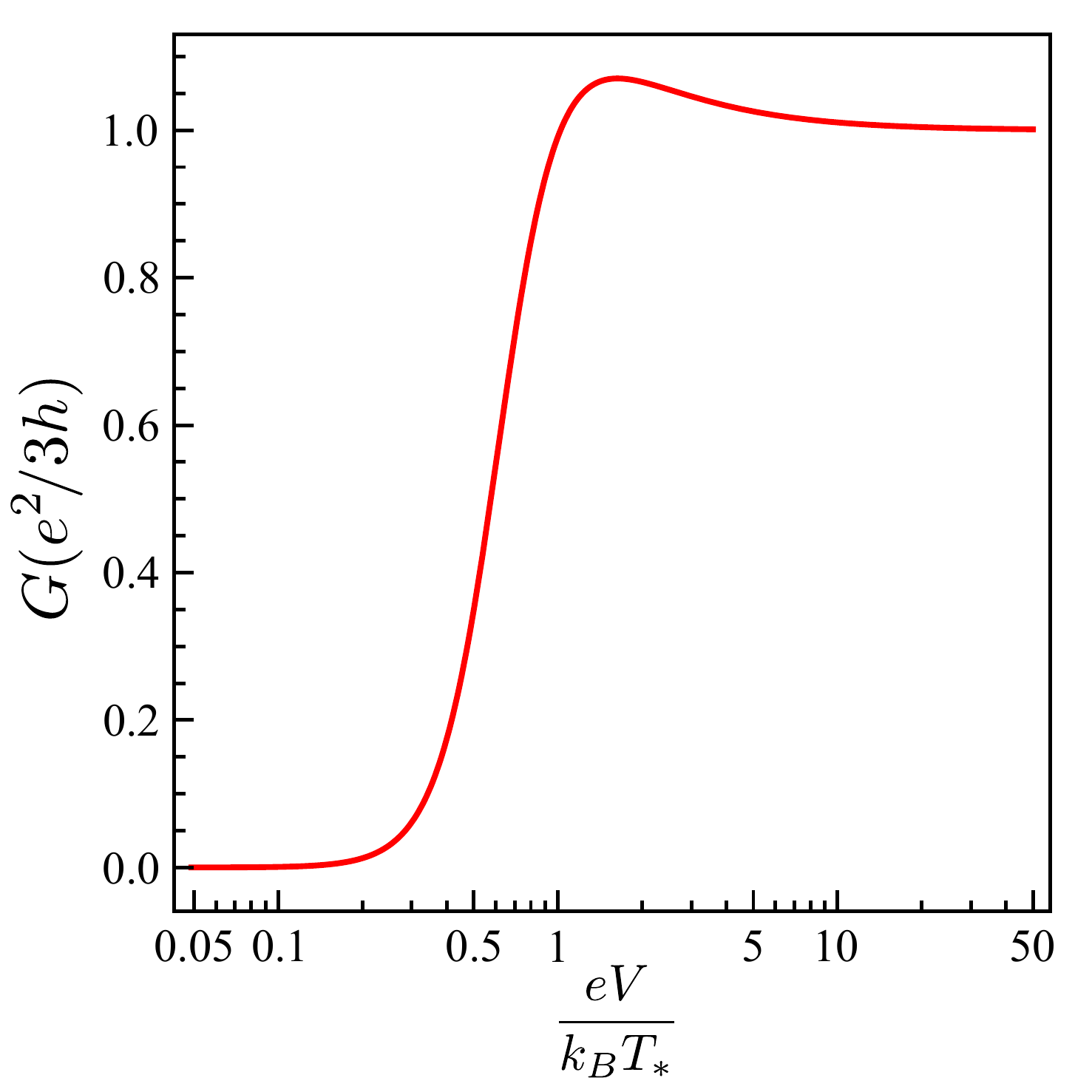}
\caption{The universal zero-temperature differential conductance (in the unit of $G_0=e^2/3h$) as a function of $eV/k_B T_*$.}\label{ucurvev}
\end{center}
\end{figure}

In the main text, we showed that the two-sites quantum island in quasi-ballistic regime falls into the same universality class as the weak tunneling between fractional quantum Hall edge states at filling $\nu=1/3$. The latter problems are described by a boundary sine-Gordon model whose Bethe ansatz solution for the charge current has been thoroughly investigated~\cite{Kane_1992, Chamon_1995, Fendley_PRL_1995, Fendley_PRB_1995}. In particular, 
the Bethe Ansatz equations can be solved analytically at vanishing temperature by using the Wiener-Hopf technique~\cite{Fendley_PRB_1995}, providing a closed series representation of the universal scaling function for the current in the low and large bias voltage regimes at zero temperature:
\begin{equation}\nonumber
I=\begin{cases}
\frac{e^2V}{h}\sum\limits_{n=1}^\infty (-1)^{n+1}\frac{\sqrt{\pi}\Gamma\left(3n\right)}{2\Gamma(n)\Gamma\left(\frac{3}{2}+2n\right)}\;x^{4n},&x<x_0\\
\frac{e^2V}{3h}\Bigg[1{-}\sum\limits_{n=1}^\infty ({-}1)^{n+1}\frac{\sqrt{\pi}\Gamma\left(\frac{n}{3}\right)}{6\Gamma(n)\Gamma\left(\frac{3}{2}{-}\frac{2n}{3}\right)}x^{-\frac{4n}{3}}\Bigg],\!&x>x_0,
\end{cases}
\end{equation}
where $x=eV/k_B T'_B$ with $T'_B$ being the boundary temperature and the convergence radius of the series expansion reads $x_0=\sqrt{2}/3^{3/4}$. We then matched the above expression of current for $x>x_0$ stopping the series at $n=1$ with that obtained perturbatively Eq.~\eqref{current2} providing the matching $T'_B=T_*$. The resulting zero temperature differential conductance is presented in Eq.~\eqref{diffV} and corresponding universal curve is depicted in Fig.~\ref{ucurvev}.

At finite temperature, the current can be computed in a standard way~\cite{Fendley_PRL_1995,Fendley_PRB_1995} from the solution of the so-called Thermodynamical Bethe Ansatz (TBA) equations, an approach that  we summarize here for completeness. The sine-Gordon model at $\nu=1/3$
has a spectrum consisting in a pair of kink-antikink, carrying charge $q=\pm e$ in our case, and of a breather which can be viewed as a kink-antikink neutral boundstate. The  corresponding quasiparticle modes are denoted $A_{a}(\theta)$, with the quantum number $a = +,-,0$ labeling the kink, the antikink, and the breather respectively. Momentum $p$ is parametrized by a rapidity $\theta=\ln \frac{p v_{\rm F}}{k_BT}$.
Those quasiparticles are not free: while integrability results in the many-body scattering between quasiparticles factorizing as elementary two-body scattering, just as in a free theory, the interaction between two quasiparticles $A_{a_1}(\theta_1)$ and 
$A_{a_2}(\theta_2)$
is encoded in a non-trivial two-body scattering matrix $S_{a_1a_2}(\theta_1-\theta_2)$~\cite{Zamolo2_79}.

As a result of this interaction, at finite temperature the quasiparticles distributions are not that of free particles, but rather obey a set of non-linear equations.
In the thermodynamical limit $L\to\infty$ (with $L$ the system size), we write the densities of occupied quasiparticles as $\frac{k_B T L}{2\pi\hbar v_{\rm F}}\; \rho_a(\theta)$ with $ \rho_a$ the reduced densities. The reduced densities can then be encoded by pseudo energies $\epsilon_c=\epsilon_\pm$ and $\epsilon_0$ via $\rho_a = f_a\;P_a$ where $P_a = \partial_\theta\epsilon_a$ is the total (occupied+empty) density of quasiparticles, and $f_a=\big(1+e^{\epsilon_a-\mu_a}\big)^{-1}$ is the occupation function. We also introduce the functions $L_a(\theta)=\ln\big( 1+e^{\epsilon_a(\theta)-\mu_a}  \big)$, which are connected to the densities via $\partial_\theta L_a = P_a-\rho_a$. The reduced chemical potentials read: 
$$\mu_0 = 0\quad;\quad \mu_\pm = \pm\frac{eV}{2k_BT}.$$ 

The TBA equations for the pseudoenergies  read
\begin{eqnarray}
\epsilon_c(\theta) &=& \frac1{\pi\cosh(2\theta)}\star L_0(\theta),
\label{tba0}\\
\epsilon_0(\theta) &=& \frac1{\pi\cosh(2\theta)}\star \Big(L_+(\theta) + L_-(\theta)\Big),
\label{tbakinks}
\end{eqnarray}
where the convolution is defined by $f(\theta)\star g(\theta) = \int d\theta'\,f(\theta')g(\theta-\theta')$.
Equations~(\ref{tba0},\ref{tbakinks})
have to be supplemented with the boundary conditions $\epsilon_c(\theta)
\underset{\theta\gg1}{\simeq} e^\theta$ 
and 
$\epsilon_0(\theta)
\underset{\theta\gg1}{\simeq} \sqrt2\;e^\theta$.
These equations are solved numerically, yielding the densities $\rho_a$. The current is then obtained through a rate equation~\cite{Fendley_PRL_1995,Fendley_PRB_1995}
\begin{equation}
I=\frac{ e k_\textsc{b} T}{2\pi\hbar}\int d\theta (\rho_+-\rho_-)(\theta) {\cal T}(\theta),
\label{currentTBA}
\end{equation}
where ${\cal T}(\theta)$ is the probability that a kink is scattered into an anti-kink at the impurity site.
This probability depends on the boundary coupling, and reads
${\cal T}(\theta) = \frac{(T/T_\textsc{b})^4}{e^{4\theta}+(T/T_\textsc{b})^4}$ where $T_\textsc{b}$ is the scale at which ${\cal T} =\frac12$. It is related to the scale $T_*$ as $T_\textsc{b} =\frac{\Gamma(\frac14)}{6\sqrt\pi\Gamma(\frac34)}\; T_*$~\cite{Fendley_PRB_1995}. We then express the linear conductance $G(V=0,T) =\lim_{V\to 0}\frac{\partial I}{\partial V}$ as
\begin{equation}
G(V\!=\!0,T)=\frac{e^2}{h}\!\int\! \frac{d\theta}{1+e^{\epsilon_c(\theta)}}\;\frac1{\cosh^2(2\theta+2\ln\frac T{T_\textsc{b}})}.
\label{condG0}
\end{equation}
The exact low- and high-temperature asymptotic behaviors of Eq.~\eqref{condG0} can be derived analytically using Keldysh perturbation theory~\cite{Boulat2019} and are presented in Eq.~\eqref{asymptotesG}. The full conductance curve together with its asymptotic behaviors 
 is depicted in Fig.~\ref{ucurvet}.

\bibliography{biblio.bib}

\end{document}